\documentclass[12pt]{article}

\title{A stochastic Keller-Segel model of chemotaxis
}

\usepackage{amsthm,amsmath,amssymb,a4wide,epsf,color}

cm \hsize=18 true cm \topmargin=-2cm \oddsidemargin=-0.5cm
\def\mb#1{\setbox0=\hbox{$#1$}\kern-.025em\copy0\kern-\wd0
\kern-0.05em\copy0\kern-\wd0\kern-.025em\raise.0233em\box0}

\begin{document}

\author{Pierre-Henri Chavanis}
\maketitle
\begin{center}
Laboratoire de Physique Th\'eorique (CNRS UMR 5152), \\
Universit\'e
Paul Sabatier,\\ 118, route de Narbonne, 31062 Toulouse Cedex 4, France\\
E-mail: {\it chavanis{@}irsamc.ups-tlse.fr\\
 }
\vspace{0.5cm}
\end{center}

\begin{abstract}

We introduce stochastic models of chemotaxis generalizing the
deterministic Keller-Segel model. These models include fluctuations
which are important in systems with small particle numbers or close to
a critical point. Following Dean's approach, we derive the exact
kinetic equation satisfied by the density distribution of cells. In
the mean field limit where statistical correlations between cells are
neglected, we recover the Keller-Segel model governing the smooth
density field. We also consider hydrodynamic and kinetic models of
chemotaxis that take into account the inertia of the particles and
lead to a delay in the adjustment of the velocity of cells with the
chemotactic gradient. We make the connection with the Cattaneo model of 
chemotaxis and the telegraph equation.

\end{abstract}

\maketitle

\section{Introduction}
\label{sec_introduction}

In biology, many organisms (bacteria, amoebae, cells,...) or social
insects (like ants, swarms,...) interact through the process of chemotaxis
\cite{armitage,kessin,murray}. Chemotaxis is a long-range interaction
that accounts for the orientation of individuals along chemical
signals that they produce themselves. Famous examples of biological
species experiencing chemotaxis are the slime mold amoebae {\it
Dictyostelium discoideum}, the flagellated bacteria {\it Salmonella
typhimurium} and {\it Escherichia coli}, the human endothelial cells
etc. When the interaction is attractive, chemotaxis is responsible for
the self-organization of the system into coherent structures such as
peaks, clusters, aggregates, fruiting bodies, periodic patterns,
spirals, rings, spots, honeycomb patterns, stripes or even filaments.
This spontaneous organization has been observed in several experiments
\cite{bonner,ford1,ford2,berg1,berg2,woodward,gazit,levitov,tsimring,carmeliet,firtel,embo,szabo}  and numerical simulations
\cite{schweitzer,sg,ben,tyson1,tyson2,brenner,hp,ph,sc,dh,gamba,post,filbet,sopik,virial1,csbio,holm}. Chemotactic
attraction is therefore a leading mechanism to account for the
morphogenesis and self-organization of biological systems. For
example, it has been advocated to explain aggregation patterns in
bacteria, tissue organization during embryonic growth, cell guidance,
fish skin pigmentation patterning, angiogenesis in tumour progression
and wound healing, formation of plaques in Alzheimer's disease,
dynamics of blood vessel formation etc \cite{ph,grima}.  It is
fascinating to realize that the self-organization of chemotactic
species in biology shares some analogies with the self-organization of
galaxies in astrophysics and large-scale vortices (like Jupiter's
great red spot) in two-dimensional turbulence \footnote{These
analogies are intrinsically due to the long-range attractive nature of
the interaction. In particular, self-gravitating systems, 2D vortices
and chemotactic species interact through a field produced by the
distribution of particles via a Poisson equation (or its
generalizations). Furthermore, the process of self-organization is
described by relatively similar relaxation equations corresponding to
nonlinear mean field Fokker-Planck equations \cite{gfp}. Therefore,
self-gravitating systems, 2D vortices and chemotactic species share
many analogies despite their very different physical nature. These
striking analogies have been emphasized by the author in several
papers
\cite{csr,houches,crrs,cras,degrad,gfp}.}.  A first successful model of
chemotactic aggregation is provided by the Keller-Segel (KS) model \cite{ks}
introduced in 1970. The standard KS model can be written as
\begin{eqnarray}
\label{intro1}
\frac{\partial\rho}{\partial t}=\nabla\cdot (D_*\nabla\rho-\chi\rho\nabla c),
\end{eqnarray}
\begin{eqnarray}
\label{intro2}
\frac{\partial c}{\partial t}=D_c\Delta c-k c+h \rho.
\end{eqnarray}
It consists in two coupled differential equations that govern the
evolution of the density of cells (or other biological entities)
$\rho({\bf r},t)$ and the evolution of the secreted chemical $c({\bf
r},t)$. The first equation (\ref{intro1}) is a drift-diffusion
equation. The cells diffuse with a diffusion coefficient $D_{*}$ and
they also move in a direction of a gradient of the chemical
(chemotactic drift). The chemotactic sensitivity $\chi$ is a measure
of the strength of the influence of the chemical gradient on the flow
of cells. The coefficient $\chi$ can be positive or negative. In
the first case (chemoattraction), the particles climb the chemical
gradient and form clusters. In the second case (chemorepulsion), they
descend the chemical gradient and repell each other. In that case, the
chemical acts like a poison. The second equation (\ref{intro2}) in the
KS model is a reaction-diffusion equation. The chemical is produced by
the bacteria with a rate $h$ and is degraded with a rate $k$. It also
diffuses with a diffusion coefficient $D_{c}$. When chemotactic
attraction prevails over diffusion, the KS model describes a {\it
chemotactic collapse} leading to aggregates or Dirac peaks. There is a
vast literature on this subject. We refer to Perthame
\cite{perthame} for numerous references in applied mathematics and to
Chavanis \cite{mass} for additional references in physics.

The first equation of the KS model can be interpeted as a mean-field
Smoluchowski equation describing a system of Brownian particles in
interaction. On the other hand, in the limit of large diffusivity of
the chemical, we can make a quasi-stationary approximation $\partial
c/\partial t\simeq 0$ in the second equation and obtain the screened
Poisson equation. We are led therefore to the simplified Keller-Segel
model
\begin{eqnarray}
\label{intro3}
\frac{\partial\rho}{\partial t}=\nabla\cdot (D_*\nabla\rho-\chi\rho\nabla c),
\end{eqnarray}
\begin{eqnarray}
\label{intro4}
\Delta c-k_{0}^{2}c=-\lambda\rho,
\end{eqnarray}
where we have set $k_{0}^2=k/D_c$ and $\lambda=h/D_c$. In the absence
of degradation of the chemical ($k_{0}=0$), the field equation (\ref{intro4})
reduces to the Poisson equation $\Delta c=-\lambda\rho$ (see
\cite{jager} and Appendix C of \cite{csbio} for a precise
justification of these approximations). In that case, the Keller-Segel
(KS) model becomes isomorphic to the Smoluchowski-Poisson (SP) system
\begin{eqnarray}
\label{intro5}
\frac{\partial\rho}{\partial t}
=\nabla\cdot \left\lbrack \frac{1}{\xi}\left (\frac{k_{B}T}{m}\nabla\rho+\rho\nabla\Phi\right )\right\rbrack,
\end{eqnarray}
\begin{eqnarray}
\label{intro6}
\Delta\Phi=S_d G\rho,
\end{eqnarray}
describing a system of overdamped self-gravitating Brownian particles
in the mean field approximation
\cite{crs,sc,post,sopik,tcoll,virial1,virial2,mass,acedo,exact}. We have the
correspondances: $D_{*}=k_{B}T/\xi m$, $\chi=1/\xi$, $c=-\Phi$,
$\lambda=S_{d}G$. {\it In particular, the concentration of the secreted
chemical $c({\bf r},t)=-\Phi({\bf r},t)$ in biology plays the role of
the gravitational potential (with the opposite sign) in astrophysics} \footnote{One great achievement of Keller \& Segel
\cite{ks} was to interpret slime mold aggregation as a manifestation
of a fundamental instability in a uniform distribution of amoebae and
acrasin (chemoattractant).  As noticed in \cite{jeansbio1,jeansbio2},
this instability is closely related to the Jeans gravitational
instability in astrophysics \cite{jeans}.}.  More generally, when we
consider a system of Brownian particles interacting via an arbitrary  binary
potential $u({\bf r}-{\bf r}')$ and make a mean-field approximation
\cite{martzel,hb1,hb2}, we obtain the mean-field Smoluchowski equation
\begin{eqnarray}
\label{intro7}
\frac{\partial\rho}{\partial t}
=\nabla\cdot \left\lbrack \frac{1}{\xi}\left (\frac{k_{B}T}{m}\nabla\rho+\rho\nabla\Phi\right )\right\rbrack,
\end{eqnarray}
\begin{eqnarray}
\label{intro8}
\Phi({\bf r},t)= \int  \rho({\bf r}',t)u({\bf r}-{\bf r}')\, d{\bf r}'.
\end{eqnarray}
The main difference between models (\ref{intro1})-(\ref{intro2}) and
(\ref{intro7})-(\ref{intro8}) comes from the equation for the field
$c({\bf r},t)$ or $\Phi({\bf r},t)$. Equation (\ref{intro2}) is
non-markovian since the concentration of the chemical $c({\bf r},t)$
at time $t$ depends on the concentration of the bacteria and of the
chemical at earlier times. By contrast, Eq.  (\ref{intro8}) is
markovian since the potential $\Phi({\bf r},t)$ is assumed to be
instantaneously produced by the distribution of particles.

It is important to note that the Keller-Segel model is a {\it mean
field} model which ignores fluctuations. This implicitly assumes that
the number of cells $N\rightarrow +\infty$ and that we are far from a
critical point \cite{hb5}. Now, in biology, the number of particles in
the system can be relatively small. Furthermore, from the statistical
physics viewpoint, it is natural to investigate the role of
fluctuations during chemotaxis. In order to go beyond the mean field
approximation, some authors
\cite{schweitzer,stevens,ng,csbio} have proposed to return to a
corpuscular description of the dynamics and to describe the motion of
the particles (chemotactic species or ``active'' walkers) by $N$
coupled stochastic Langevin equations of the form
\begin{eqnarray}
\label{intro9}
\frac{d{\bf r}_{i}}{dt}=\chi\nabla c_{d}({\bf r}_{i}(t),t)+\sqrt{2D_{*}}{\bf R}_{i}(t),
\end{eqnarray}
\begin{eqnarray}
\label{intro10}
\frac{\partial c_{d}}{\partial t}=D_{c}\Delta c_{d}-kc_{d}+h\sum_{i=1}^{N}\delta({\bf r}-{\bf r}_{i}(t)),
\end{eqnarray}
where ${\bf r}_{i}(t)$ denote the positions of the
particles, $c_{d}({\bf r},t)$ is the exact field of secreted chemical
and ${\bf R}_{i}(t)$ is a white noise satisfying $\langle {\bf
R}_{i}(t)\rangle ={\bf 0}$ and $\langle
R_{i,\alpha}(t)R_{j,\beta}(t')\rangle=\delta_{ij}\delta_{\alpha\beta}\delta
(t-t')$ where $i=1,...,N$ refer the the particles and $\alpha=1,...,d$
to the dimensions of space. Note that the motion of cells is treated
on an individual basis but the chemical signals are treated in the
continuum limit. This separation of scales appears to be reasonable in
most applications. In the mean field approximation, these stochastic
equations lead to the KS model
(\ref{intro1})-(\ref{intro2})\footnote{Stevens \cite{stevens} gives
the first rigorous derivation (in the mathematical sense) of the KS
model from an interacting stochastic many-particle system where the
interaction between the particles is rescaled in a moderate way as the
population size $N$ tends to infinity.}. When the reaction-diffusion
equation (\ref{intro10}) is replaced by a Markovian equation of the
form
\begin{eqnarray}
\label{intro11}
\Delta c_{d}-k_{0}^{2}c_{d}=-\lambda \sum_{i=1}^{N}\delta({\bf r}-{\bf r}_{i}(t)),
\end{eqnarray}
we obtain a simplified model of chemotaxis that leads to the
simplified KS model (\ref{intro3})-(\ref{intro4}) in the mean field
approximation. More generally, for Brownian particles interacting
via a binary potential of interaction $u({\bf r}-{\bf r}')$, one
obtains the stochastic model 
\begin{eqnarray}
\label{intro12}
\frac{d{\bf r}_{i}}{dt}=-\frac{1}{\xi}\nabla \Phi_{d}({\bf r}_{i}(t),t)+\sqrt{\frac{2k_{B}T}{\xi m}}{\bf R}_{i}(t),
\end{eqnarray}
\begin{eqnarray}
\label{intro13}
\Phi_{d}({\bf r},t)= \sum_{i=1}^{N}m \ u({\bf r}-{\bf r}_{i}(t)),
\end{eqnarray}
considered in
\cite{kk,dean,mt,martzel,arb,hb1,hb2,virial2,hb5}. In the mean field approximation \cite{martzel,hb1,hb2}, these
equations yield the mean-field Smoluchowski equation
(\ref{intro7})-(\ref{intro8}).

In systems with weak long-range interactions, the mean field
approximation is expected to become exact in a proper thermodynamic
limit $N\rightarrow +\infty$ such that the strength of the potential
scales like $1/N$ while the volume $V$ remains of order unity
\cite{hb1}. In the context of chemotaxis, the differences between mean
field and non mean field models have been discussed by Grima
\cite{grima} who showed situations where the mean field approximation
fails to predict the width of the aggregate sizes. In particular, the
disagreement is very severe close to the critical point where we know
that mean field approximation breaks down in general
\cite{kadanoff}. This is because the fluctuations become very
important so that it is not possible to neglect the two-body
correlation function anymore \cite{hb5}. On the other hand, the mean
field approximation assumes that the number of particles $N\gg 1$. In
stellar systems and plasmas, this is always the case. However, for
biological systems, the number of interacting bacteria or cells is
frequently less than a few thousands so that finite $N$ effects and
statistical fluctuations are important.  In view of these remarks, it
is highly desirable to obtain stochastic kinetic equations that take
into account fluctuations and that go beyond the deterministic mean
field Keller-Segel model. Such equations are discussed in the present
paper. In the first part of the paper (Sec. \ref{sec_sks}), following
Dean's approach \cite{dean}, we derive the exact kinetic equation
satisfied by the density distribution of chemotactic species. This
equation takes into account stochastic fluctuations and memory effects
present in the field equation for the secreted chemical.  If we
average over the noise, we recover the hierarchy of kinetic equations
discussed by Newman \& Grima \cite{ng}. If we make a mean-field
approximation, we recover the Keller-Segel model \cite{ks}. Therefore,
this exact stochastic kinetic equation generalizes several models
introduced in the chemotactic literature. We also propose a simplified
kinetic equation for a coarse-grained density field (instead of a sum
of $\delta$-functions) keeping track of fluctuations. This equation
(\ref{sks14})-(\ref{sks15}) could be of practical interest in
chemotaxis. In the second part of the paper (Secs. \ref{sec_c} to
\ref{sec_sk}), we note that the Keller-Segel model is a parabolic
model which neglects the inertia of the particles and which assumes an
instantaneous adjustment of the velocity with the chemotactic
gradient. We consider hyperbolic models that generalize this parabolic
model. We first consider the Cattaneo model of chemotaxis \cite{dh}
which consists in introducing a delay in the establishment of the
current (Sec. \ref{sec_c}). Then, we consider hydrodynamic models
including a friction force (Sec. \ref{sec_mm}). Using a semi-linear
approximation, we show that the Cattaneo model can be recovered from
these hydrodynamic equations \cite{hb5}. In Sec. \ref{sec_sh}, we
generalize these models so as to take into account fluctuations. This
leads to stochastic hyperbolic models of chemotaxis which generalize
the ordinary deterministic parabolic Keller-Segel model. Finally, in
Sec. \ref{sec_sk}, we develop a kinetic theory of chemotactic species
in phase space taking into account the inertia of the particles and
the discrete nature of the system. We derive stochastic kinetic
equations that should improve the description of the cells'
motion. The link with the parabolic and hyperbolic models is also
discussed.

This paper adapts the results of \cite{hb5} to the context of
chemotaxis with complements and amplification. Although these
different stochastic equations (in particular the parabolic ones) are
well-known in statistical physics
\cite{kk,dean,mt,arb,hb5} their application to the context of
chemotaxis, proposed in \cite{hb5}, is new and is an important
contribution of the present paper.

\section{The stochastic Keller-Segel model}
\label{sec_sks}

In this section, we introduce a stochastic model of chemotaxis,
generalizing the Keller-Segel model, by taking into account
fluctuations. Let us first derive the exact kinetic equation satisfied
by the density distribution of cells whose dynamics is described by
the coupled stochastic Langevin equations
(\ref{intro9})-(\ref{intro10}). We follow Dean's approach
\cite{dean}. The exact density field, expressed in terms of
$\delta$-functions, can be written
\begin{eqnarray}
\label{sks1}
\rho_{d}({\bf r},t)=\sum_{i=1}^{N}\rho_{i}({\bf r},t)=\sum_{i=1}^{N}\delta({\bf r}-{\bf r}_{i}(t)).
\end{eqnarray}
For any function $F({\bf r})$, we have $F({\bf r}_{i}(t))=\int \rho_{i}({\bf r},t)F({\bf r})d{\bf r}$. Now, using Ito's calculus \cite{oksendal}, one has
\begin{eqnarray}
\label{d1}
\frac{dF({\bf r}_{i})}{dt}=\int \rho_{i}({\bf r},t)\left\lbrack \chi \nabla F({\bf r})\cdot\nabla c_{d}({\bf r},t)+\sqrt{2D_{*}}\nabla F({\bf r})\cdot {\bf R}_{i}(t)+D_{*}\Delta F({\bf r})\right \rbrack \, d{\bf r}.
\end{eqnarray}
Integrating by parts, we obtain
\begin{eqnarray}
\label{d2}
\frac{dF({\bf r}_{i})}{dt}=\int F({\bf r})\left\lbrack -\chi \nabla \cdot (\rho_{i}({\bf r},t)\nabla c_{d}({\bf r},t))-\sqrt{2D_{*}}\nabla \cdot (\rho_{i}({\bf r},t){\bf R}_{i}(t))+D_{*}\Delta \rho_{i}({\bf r},t)\right \rbrack \, d{\bf r}.
\end{eqnarray}
Then, using $dF({\bf r}_i)/dt=\int \partial_t\rho_{i}({\bf r},t)F({\bf r}) d{\bf r}$ and comparing with Eq. (\ref{d2}), we get (using the fact that $F$ is an arbitrary function)
\begin{eqnarray}
\label{d3}
\frac{\partial\rho_{i}}{\partial t}=-\chi \nabla \cdot (\rho_{i}({\bf r},t)\nabla c_{d}({\bf r},t))-\sqrt{2D_{*}}\nabla \cdot (\rho_{i}({\bf r},t){\bf R}_{i}(t))+D_{*}\Delta \rho_{i}({\bf r},t).
\end{eqnarray}
Summing this relation over the $i$, we finally obtain
\begin{eqnarray}
\label{d4}
\frac{\partial\rho_d}{\partial t}({\bf r},t)=D_{*}\Delta \rho_{d}({\bf r},t)-\chi \nabla \cdot (\rho_{d}({\bf r},t)\nabla c_{d}({\bf r},t))-\sqrt{2D_{*}}\sum_{i=1}^{N}\nabla \cdot (\rho_{i}({\bf r},t){\bf R}_{i}(t)).
\end{eqnarray}
Now, the last term can be rewritten \cite{dean}:
\begin{eqnarray}
\label{d5}
-\sum_{i=1}^{N}\nabla \cdot (\rho_{i}({\bf r},t){\bf
R}_{i}(t))=\nabla\cdot (\rho_{d}^{1/2}({\bf r},t){\bf R}({\bf r},t)),
\end{eqnarray} 
where ${\bf R}({\bf r},t)$ is a Gaussian random field such that
$\langle {\bf R}({\bf r},t)\rangle ={\bf 0}$ and $\langle
R_{\alpha}({\bf r},t)R_{\beta}({\bf r}',t')\rangle
=\delta_{\alpha\beta}\delta({\bf r}-{\bf r}')\delta(t-t')$.
Therefore, the system of equations satisfied by the exact density field expressed
in terms of $\delta$-functions is
\begin{eqnarray}
\label{sks5}
\frac{\partial \rho_{d}}{\partial t}({\bf r},t)=D_{*}\Delta\rho_{d}({\bf r},t)-\chi\nabla\cdot (\rho_{d}({\bf r},t)\nabla c_{d}({\bf r},t))+\nabla \cdot \left (\sqrt{2D_{*}\rho_{d}({\bf r},t)}{\bf R}({\bf r},t)\right ),
\end{eqnarray}
\begin{eqnarray}
\label{sks5b}
\frac{\partial c_{d}}{\partial t}=D_{c}\Delta c_{d}({\bf r},t)-kc_{d}({\bf r},t)+h\rho_{d}({\bf r},t).
\end{eqnarray}
The first and third terms in the r.h.s. of Eq. (\ref{sks5}) correspond
to a pure Brownian motion and the second term takes into account
chemotaxis, i.e the attraction or repulsion of the cells by the
chemical. As noted by Dean \cite{dean}, the noise in Eq. (\ref{sks5})
appears not additively but multiplicatively.

Integrating Eq. (\ref{intro10}), the concentration of the chemical can be
expressed in terms of the cell paths as  \cite{ng}:
\begin{eqnarray}
\label{sks6}
c_{d}({\bf r},t)=h\int d{\bf r}'\int_{0}^{t}dt' G({\bf r}-{\bf r}',t-t')\sum_{i=1}^{N}\delta({\bf r}'-{\bf r}_{i}(t')),
\end{eqnarray}
where the Green function for the chemical diffusion equation is given by
\begin{eqnarray}
\label{sks7}
G({\bf r},t)=(4\pi D_{c}t)^{-d/2}{\rm exp}\left\lbrack -\frac{r^{2}}{4D_{c}t}-kt\right\rbrack.
\end{eqnarray}
The gradient of the concentration field is 
\begin{eqnarray}
\label{sks8}
\nabla c_{d}({\bf r},t)=h\int d{\bf r}'\int_{0}^{t}dt' \nabla G({\bf r}-{\bf r}',t-t')\rho_{d}({\bf r}',t').
\end{eqnarray}
Substituting Eq. (\ref{sks8}) in Eq. (\ref{sks5}),
we obtain
\begin{eqnarray}
\label{sks9}
\frac{\partial \rho_{d}}{\partial t}({\bf r},t)=D_{*}\Delta\rho_{d}({\bf r},t)-\chi h \nabla\cdot \left\lbrack \rho_{d}({\bf r},t)\nabla\int d{\bf r}'\int_{0}^{t}dt'  G({\bf r}-{\bf r}',t-t')\rho_{d}({\bf r}',t') \right\rbrack\nonumber\\
+\nabla \cdot \left (\sqrt{2D_{*}\rho_{d}({\bf r},t)}{\bf R}({\bf r},t)\right ).
\end{eqnarray}
If we average over the noise and introduce the
smooth density $\rho({\bf r},t)=\langle
\rho_{d}({\bf r},t)\rangle$, we recover Eq. (9) of Newman \& Grima
\cite{ng}:
\begin{eqnarray}
\label{sks10}
\frac{\partial \rho}{\partial t}({\bf r},t)=D_{*}\Delta\rho({\bf r},t)-\chi h \nabla\cdot  \int d{\bf r}'\int_{0}^{t}dt'  \lbrack \nabla G({\bf r}-{\bf r}',t-t')\rbrack \langle \rho_{d}({\bf r},t)\rho_{d}({\bf r}',t')\rangle.
\end{eqnarray}
If we make a mean field approximation $\langle \rho_{d}({\bf
r},t)\rho_{d}({\bf r}',t')\rangle\simeq \rho({\bf r},t) \rho({\bf
r}',t')$ in Eq. (\ref{sks10}), we recover the Keller-Segel model
\cite{ks}:
\begin{eqnarray}
\label{sks11}
\frac{\partial \rho}{\partial t}({\bf r},t)=D_{*}\Delta\rho({\bf r},t)-\chi  \nabla\cdot (\rho({\bf r},t)\nabla c({\bf r},t)), 
\end{eqnarray}
with
\begin{eqnarray}
\label{sks12}
c({\bf r},t)=h \int d{\bf r}'\int_{0}^{t}dt'  G({\bf r}-{\bf r}',t-t') \rho({\bf r}',t').
\end{eqnarray}
Given the definition of the Green function $G$, the smooth concentration $c({\bf r},t)$ is solution of the reaction-diffusion equation
\begin{eqnarray}
\label{sks13}
\frac{\partial c}{\partial t}=D_{c}\Delta c-kc+h\rho.
\end{eqnarray}
We also note, for future reference, that the steady solutions of the
KS model (\ref{sks11}) correspond to a mean field Boltzmann-like distribution
\begin{eqnarray}
\label{sks13b}
\rho=A e^{c/T_{eff}},
\end{eqnarray}
where $T_{eff}=D_{*}/\chi$ is an effective temperature given by an Einstein
relation.

Grima \cite{grima} has shown that the mean field approximation may
lead to wrong results if we are close to a critical point or if the
number of particles is not large enough. Therefore, it may be useful
to have a more general model than the Keller-Segel model
(\ref{intro1})-(\ref{intro2}) which keeps track of
fluctuations. Equations (\ref{sks5})-(\ref{sks5b}) are exact and
contain the same information as the $N$-body stochastic
Langevin equations (\ref{intro9})-(\ref{intro10}). They are not very useful for
practical purposes since they govern the evolution of a density field
which is expressed as a sum of $\delta$-functions. It is easier to
directly solve the equivalent $N$-body stochastic Langevin equations
(\ref{intro9})-(\ref{intro10}). However, using phenomenological
arguments like those described in \cite{arb,hb5}, we can consider a
spatio-temporal coarse-grained distribution $\overline{\rho}({\bf
r},t)$ which smoothes out the exact density field $\rho_{d}({\bf
r},t)$ while keeping track of fluctuations. We also assume that the
spatio-temporal window is sufficiently small so that we can make the
approximation $\overline{\rho}^{(2)}({\bf r},{\bf r}',t)\simeq
\overline{\rho}({\bf r},t)\overline{\rho}({\bf r}',t)$. In that case,
we obtain the stochastic Keller-Segel model for the coarse-grained
distribution
\begin{eqnarray}
\label{sks14}
\frac{\partial \overline{\rho}}{\partial t}({\bf r},t)=D_{*}\Delta\overline{\rho}({\bf r},t)-\chi  \nabla\cdot (\overline{\rho}({\bf r},t)\nabla \overline{c}({\bf r},t))
+\nabla \cdot \left (\sqrt{2D_{*}\overline{\rho}({\bf r},t)}{\bf R}({\bf r},t)\right ),
\end{eqnarray}
\begin{eqnarray}
\label{sks15}
\frac{\partial \overline{c}}{\partial t}=D_{c}\Delta \overline{c}-k\overline{c}+h\overline{\rho},
\end{eqnarray}
generalizing the deterministic Keller-Segel model
(\ref{intro1})-(\ref{intro2}). This equation is one of the most
important result of this paper. As shown in Appendix B of
\cite{hb5}, the form of
the noise term in Eq. (\ref{sks14}) can be obtained from the
general theory of fluctuations developed in Landau \& Lifshitz
\cite{ll}. This provides another, direct, justification of the 
stochastic Eq. (\ref{sks14}). As shown in \cite{hb5}, the mean field
approximation breaks down close to a critical point because the
two-body correlation function diverges. In that case, it may be more
relevant to use the stochastic Keller-Segel model
(\ref{sks14})-(\ref{sks15}) including fluctuations instead of the
deterministic Keller-Segel model (\ref{intro1})-(\ref{intro2}).

It is also very important to take into account fluctuations when the
system can be found in several metastable states. If we introduce the
coarse-grained free energy functional
\begin{eqnarray}
\label{cg5}
F_{c.g.}[\overline{\rho},\overline{c}]=\frac{D_{*}}{\chi}\int \overline{\rho}\ln \overline{\rho} \, d{\bf r}+\frac{1}{2h}\int \left\lbrack D_{c}(\nabla \overline{c})^{2}+k\overline{c}^{2}\right \rbrack \, d{\bf r}-\int \overline{\rho} \ \overline{c}\, d{\bf r},
\end{eqnarray}
we can write the stochastic equation (\ref{sks14}) in the form
\begin{eqnarray}
\label{cg6}
\frac{\partial\overline{\rho}}{\partial t}=\nabla\cdot \left\lbrack \chi\overline{\rho}({\bf r},t)\nabla\frac{\delta F_{c.g.}}{\delta\overline{\rho}}\right\rbrack+\nabla\cdot \left (\sqrt{2D_{*}\overline{\rho}({\bf r},t)}{\bf R}({\bf r},t)\right ).
\end{eqnarray}
This equation can be viewed as a Langevin equation for the field
$\overline{\rho}({\bf r},t)$. The evolution of the probability of the
density distribution $W[\overline{\rho},t]$ is governed by a
Fokker-Planck equation of the form 
\begin{eqnarray}
\label{ex7add}
\frac{\partial W[\overline{\rho},t]}{\partial t}=-\int \frac{\delta}{\delta\overline{\rho}({\bf r},t)}\left\lbrace \nabla\cdot \overline{\rho}({\bf r},t)\nabla\left\lbrack D_{*}\frac{\delta}{\delta\overline{\rho}}+\chi\frac{\delta F_{c.g.}}{\delta\overline{\rho}}\right \rbrack W[\overline{\rho},t]\right\rbrace d{\bf r}.
\end{eqnarray}   
At equilibrium, we have $W[\overline{\rho}]\propto
e^{-F_{c.g.}[\overline{\rho}]/T_{eff}-\alpha\int \overline{\rho}d{\bf
r}}$ with
$F_{c.g.}[\overline{\rho}]=\frac{D_{*}}{\chi}\int\overline{\rho}\ln\overline{\rho}\,
d{\bf r}-\frac{1}{2}\int \overline{\rho} \ \overline{c}\, d{\bf r}$
(we have substituted Eq. (\ref{sks15}) with
$\partial\overline{c}/\partial t=0$ in Eq. (\ref{cg5})). For
$N\rightarrow +\infty$, the equilibrium distribution
$W[\overline{\rho}]$ is strongly peaked around the {\it global}
minimum of $F_{c.g.}[\overline{\rho}]$ at fixed mass
$M=\int\overline{\rho}\, d{\bf r}$. However, the system can remain
trapped in a metastable state (local minimum of
$F_{c.g.}[\overline{\rho}]$) for a very long time which becomes
infinite at the thermodynamic limit $N\rightarrow +\infty$. Let us be
more precise. If we ignore the noise term, Eq. (\ref{cg6}) reduces to
\begin{eqnarray}
\label{cg6n}
\frac{\partial\overline{\rho}}{\partial t}=\nabla\cdot \left\lbrack \chi\overline{\rho}({\bf r},t)\nabla\frac{\delta F_{c.g.}}{\delta\overline{\rho}}\right\rbrack,
\end{eqnarray}
which is the deterministic Keller-Segel model (\ref{sks11}). This
equation satisfies an H-theorem
\begin{eqnarray}
\label{cg6h}
\dot F=-\int \frac{1}{\chi\rho} \left (D_{*}\nabla\overline{\rho}-\chi\overline{\rho}\nabla\overline{c}\right )^{2}\, d{\bf r}-\frac{1}{h}\int \left (D_{c}\Delta \overline{c}-k\overline{c}+h\overline{\rho}\right )^{2}\, d{\bf r}\le 0,
\end{eqnarray}
with $\dot F=0$ iff the distribution is given by
Eq. (\ref{sks13b}). Therefore, a steady state is stable iff it is a
(local) minimum of free energy at fixed mass. Assuming that the free
energy is bounded from below, we know from Lyapunov's direct method
that the system will relax towards a steady state that is a {minimum}
(global or local) of the free energy functional
$F_{c.g.}[\overline{\rho}]$ at fixed mass (maxima or saddle points of
free energy are linearly dynamically unstable with respect to mean
field Fokker-Planck equations
\cite{gfp}).  If the free energy admits several local minima, the
selection of the steady state will depend on a notion of {\it basin of
attraction}. Without noise, the system remains on a minimum of free
energy forever. Now, in the presence of noise, the fluctuations can
induce {\it dynamical phase transitions} from one minimum to the
other. We should therefore see the system ``jump'' between different
states. Thus, accounting correctly for fluctuations is very important
when there exists metastable states. The probability of transition
scales as $e^{-\Delta F/T_{eff}}$ where $\Delta F$ is the barrier of
free energy between two minima. Therefore, on an infinite time, the
system will explore all the minima and will spend most time in the
global minimum.  This will be the
case only if $N$ is not too large. Indeed, for systems with long-range
interactions, the barrier of free energy $\Delta F$ scales like $N$ so
that the probability of escape from a local minimum is very small and
behaves like $e^{-N}$.  Therefore, even if the global minimum is in
principle the most probable state, metastable states are highly robust
in practice since their lifetime scales like $e^{N}$. They are thus
fully relevant for $N\gg 1$: metastable states are in practice
``stable states''. These interesting features (basin of attraction,
dynamical phase transitions, metastability,...)  would be interesting
to study in more detail in the case of chemotaxis. The study
of the stochastic Keller-Segel model will be considered in future
publications.

\section{The Cattaneo model of chemotaxis}
\label{sec_c}

The general  Keller-Segel (GKS) model \cite{ks} can be written as
\begin{eqnarray}
\label{c1}
\xi\frac{\partial\rho}{\partial t}=\nabla\cdot (D_2(\rho,c)\nabla\rho-D_1(\rho,c)\nabla c),
\end{eqnarray}
\begin{eqnarray}
\label{c2}
\frac{\partial c}{\partial t}=D_c\Delta c-k(c)c+h(c)\rho,
\end{eqnarray}
where $D_{1}=D_{1}(\rho,c)$ and $D_{2}=D_{2}(\rho,c)$ can both depend
on the concentration of the cells and of the chemical. This takes into
account microscopic constraints, like close-packing effects, that can
hinder the movement of cells and lead to nonlinear diffusion and
nonlinear mobility. The GKS model (\ref{c1})-(\ref{c2}) can be viewed
as a {\it nonlinear mean field Fokker-Planck equation} associated with a
notion of effective generalized thermodynamics \cite{gfp}. The first
equation can be written in the form of a continuity equation
$\partial_{t}\rho=-\nabla\cdot {\bf J}$ with a current
\begin{eqnarray}
\label{c3}
{\bf J}=-\frac{1}{\xi} \left (D_2(\rho,c)\nabla\rho-D_1(\rho,c)\nabla c\right ).
\end{eqnarray}
It is important to note that the GKS model is a {\it parabolic} model
like the usual heat diffusion equation. Like for the Fourier law of
heat conduction, it is assumed that the current ${\bf J}$ is
instantaneously equal to the right hand side of Eq. (\ref{c3}), that
we shall call the ``chemotactic gradient'' for future reference. In the
context of heat conduction, Cattaneo \cite{cattaneo} has proposed a
modification of Fourier's law in order to describe heat propagation
with finite speed. In the context of chemotaxis, Dolak \& Hillen
\cite{dh} have introduced a Cattaneo model for chemosensitive
movement. They assume that the current is not instantaneously equal to
the chemotactic gradient but relaxes to it with a time constant
$1/\tau$. Then, the corresponding Cattaneo model for chemosensitive
movement reads
\begin{eqnarray}
\label{c4}
\frac{\partial\rho}{\partial t}+\nabla\cdot {\bf J}=0,
\end{eqnarray}
\begin{eqnarray}
\label{c5}
\tau \frac{\partial {\bf J}}{\partial t}+{\bf J}=-\frac{1}{\xi} (D_2(\rho,c)\nabla\rho-D_1(\rho,c)\nabla c).
\end{eqnarray}
Taking the time derivative of Eq. (\ref{c4}) and using Eq. (\ref{c5}), we obtain
the {\it hyperbolic} model
\begin{eqnarray}
\label{c6}
\tau\frac{\partial^{2}\rho}{\partial t^{2}}+\frac{\partial\rho}{\partial t}=\frac{1}{\xi}\nabla\cdot (D_2(\rho,c)\nabla\rho-D_1(\rho,c)\nabla c).
\end{eqnarray}
This equation, which is second order in time, is analogous to the {\it
telegraph equation} which generalizes the diffusion equation by
introducing memory effects. For $\tau=0$, we recover the GKS
model (\ref{c1})-(\ref{c2}) as a particular case.

\section{Hydrodynamic models of chemotaxis}
\label{sec_mm}

The parabolic Keller-Segel model \cite{ks} is able to reproduce the
formation of clusters (clumps) resulting from chemotactic
collapse. This can explain experiments on bacteria like {\it
Escherichia coli} or amoebae like {\it Dictyostelium discoideum}
exhibiting pointwise concentrations
\cite{bonner,firtel,woodward,levitov,ford1,ford2,berg1,berg2}. Recently, several
experiments with human endothelial cells have shown the formation of
networks that can be interpreted as the initiation of a vasculature
\cite{gazit,carmeliet,gamba,szabo,embo}.  Cells randomly spread on a gel matrix autonomously organize to form a
continuous multicellular network which can be described as a
collection of nodes connected by chords \cite{gamba}. This process
takes place during the early stages of vasculogenesis in embryo
development. These filaments are observed in the experiments of
capillary blood vessel formation.  These structures cannot be
explained by the Keller-Segel parabolic model which generically leads
to pointwise blow-up \footnote{In fact, some Keller-Segel models
including cell kinetics can, under certain conditions, give rise to
network-like patterns (see Fig. 12c of
\cite{ph}).}. In order to account for these filaments, hyperbolic models
of chemotaxis have been introduced
\cite{gamba,coniglio,crrs,filbet,jeansbio1,csbio,talia}. They have the form
of damped hydrodynamic equations \footnote{The type of hydrodynamic
equations (\ref{mm1})-(\ref{mm2}) including a long-range mean field
interaction, a density dependent pressure and a friction force were
introduced in Chavanis \cite{gen} (see also \cite{csr}) for Langevin
particles in interaction and called the {\it damped Euler
equations}. Their application to chemotaxis and gravity was
mentioned. These equations can be derived from kinetic equations
(nonlinear mean-field Fokker-Planck equations) by using a local
thermodynamic equilibrium condition (L.T.E.) to close the hierarchy of
hydrodynamic moments
\cite{gen,virial2,csbio}. However, they remain heuristic because the
L.T.E. approximation is not rigorously justified. By contrast, in the
strong friction limit $\xi\rightarrow +\infty$, we can rigorously
derive the GKS model (\ref{c1})-(\ref{c2}), also called the {\it
generalized Smoluchowski equation}, by using a Chapman-Enskog
expansion
\cite{lemou} or a method of moments \cite{banach,csbio,gfp}.  The
model considered by Gamba {\it et al.}
\cite{gamba} (see also \cite{coniglio,embo,filbet}) corresponds to
$\xi=0$ in Eq. (\ref{mm2}). It can be derived in an asymptotic limit of kinetic
equations of a different type (see \cite{filbet} and Appendix D of
\cite{csbio}). In more recent papers \cite{talia}, the aforementioned 
authors have also included a friction force in their
model.} taking into account inertial effects :
\begin{eqnarray}
\label{mm1}
\frac{\partial \rho}{\partial t}+\nabla\cdot (\rho {\bf u})=0,
\end{eqnarray}
\begin{eqnarray}
\label{mm2}
\frac{\partial}{\partial t}(\rho {\bf u})+\nabla (\rho {\bf u}\otimes {\bf u})=-D_2(\rho,c)\nabla\rho+D_{1}(\rho,c)\nabla c-\xi\rho {\bf u}.
\end{eqnarray}
Considering the momentum equation (\ref{mm2}), the inertial term
(l.h.s.) models cells directional persistence, i.e. the natural
tendency of a particle to continue in a given direction in the absence
of any interaction.  When $D_{2}(\rho,c)$ depends only on the density,
the first term on the r.h.s. can be interpreted as a barotropic
pressure force $-\nabla p(\rho)$ (see
\cite{gfp} for different examples of equations of state). The pressure
law is expected to be linear for low densities and to increase rapidly
above a certain threshold $\sim \sigma_0$ in order to describe the
fact that the cells do not interpenetrate. For example, in
\cite{gen,degrad,csbio} we proposed to take
$p(\rho)=-\sigma_{0}T_{eff}\ln(1-\rho/\sigma_0)$ which returns the
``isothermal'' equation of state $p=\rho T_{eff}$ for dilute systems
$\rho\ll\sigma_{0}$ where the motion of an individual cell is not
impeded by the other cells \cite{sc}, and which diverges when the
cells are compressed towards the maximum density $\rho\rightarrow
\sigma_{0}$. Another possible equation of state 
is the polytropic one $p(\rho)=K\rho^{\gamma}$ \cite{lang,csbio}
taking into account anomalous transport (normal transport corresponds
to the isothermal case $\gamma=1$). The chemotactic response
$D_{1}(\rho,c)$ of the bacterium to the chemical gradient (second term
in the r.h.s. of Eq. (\ref{mm2})) can also depend on $c$ and $\rho$ so
as to take into account anomalous reactivity (the normal case
corresponds to $D_{1}(\rho,c)=\rho$ but the form
$D_{1}(\rho,c)=\rho(1-\rho/\sigma_{0})$ has also been considered to
take into account volume filling effects \cite{ph,gen,degrad,gfp}).
Finally, the last term in the r.h.s. of Eq. (\ref{mm2}) is a friction
force that measures the importance of inertial effects. It
parametrizes the tendency of the organisms to continue in a given
direction. In this inertial model, the velocity of a particle takes a
{\it finite} time $\xi^{-1}$ to get aligned with the chemotactic
gradient while in the Keller-Segel model, this alignement is assumed
to be instantaneous (see below). The ``delay'' in the alignement of
the velocity with the chemotactic gradient is similar to the idea that
is at the heart of the Cattaneo model in Sec. \ref{sec_c}.

If we neglect the friction force ($\xi=0$)
we recover the model introduced by Gamba {\it et al.}
\cite{gamba}.  Alternatively, if we neglect the inertial term (l.h.s.) 
in Eq. (\ref{mm2}) and substitute the resulting expression
\cite{gen,jeansbio1,csbio}:
\begin{eqnarray}
\label{mm2b}
\rho {\bf u}=-\frac{1}{\xi} \left (D_2(\rho,c)\nabla\rho-D_1(\rho,c)\nabla c\right ),
\end{eqnarray}
in Eq. (\ref{mm1}), we recover the GKS model. This is valid in a
strong friction limit $\xi\rightarrow +\infty$. We can also obtain a
more general model taking into account some memory effects. If we
neglect only the nonlinear term $\nabla (\rho {\bf u}\otimes {\bf u})$
in Eq. (\ref{mm2}), we obtain
\begin{eqnarray}
\label{mm3}
\frac{\partial}{\partial t}(\rho {\bf u})=-D_2(\rho,c)\nabla\rho+D_{1}(\rho,c)\nabla c-\xi\rho {\bf u},
\end{eqnarray}
which is equivalent to the Cattaneo model (\ref{c5}) with $\tau=1/\xi$.  Taking the time derivative of Eq. (\ref{mm1}) and
substituting Eq. (\ref{mm3}) in the resulting expression, we obtain a
simplified hyperbolic model keeping track of memory effects
\begin{eqnarray}
\label{mm4}
\frac{\partial^{2}\rho}{\partial t^{2}}+\xi\frac{\partial\rho}{\partial t}
=\nabla\cdot  \left (D_{2}(\rho,c)\nabla\rho-D_{1}(\rho,c)\nabla c\right ).
\end{eqnarray}
This provides a new justification (see also \cite{hb5}) of the Cattaneo
model of chemotaxis from the damped hydrodynamics equation
(\ref{mm1})-(\ref{mm2}). This can be viewed as a semi-linear
hydrodynamic model since its derivation assumes that the nonlinear
term $\nabla (\rho {\bf u}\otimes {\bf u})$ in Eq. (\ref{mm2}) can be
neglected while the full nonlinearities in the r.h.s. of
Eq. (\ref{mm2}) are taken into account.

\section{Stochastic hydrodynamic models of chemotaxis}
\label{sec_sh}

In this section, we generalize the previous hydrodynamic equations in
order to take into account fluctuations. We restrict
ourselves to the standard situation where $D_{2}=\xi D_{*}$ and
$D_{1}=\rho$. The stochastic damped Euler equations generalizing Eqs. (\ref{mm1})-(\ref{mm2}) can be written
\begin{eqnarray}
\label{sh1}
\frac{\partial \rho}{\partial t}+\nabla\cdot (\rho {\bf u})=0,
\end{eqnarray}
\begin{eqnarray}
\label{sh2}
\frac{\partial}{\partial t}(\rho {\bf u})+\nabla (\rho {\bf u}\otimes {\bf u})=-\xi D_*\nabla\rho+\rho\nabla c-\xi\rho {\bf u}-\sqrt{2D_{*}\xi^{2}\rho}\ {\bf R}({\bf r},t).
\end{eqnarray}
As shown in Appendix B of \cite{hb5}, the form of the noise in these
equations can be obtained by applying the general theory of
fluctuations developed by Landau \& Lifshitz \cite{ll}.  If we neglect
the inertial term (l.h.s.) in Eq. (\ref{sh2}) and substitute the
resulting expression
\begin{eqnarray}
\label{sh3}
\rho {\bf u}=-(D_*\nabla\rho-\chi\rho\nabla c)-\sqrt{2D_{*}\rho}\ {\bf R}({\bf r},t),
\end{eqnarray}
where $\chi=1/\xi$ in Eq. (\ref{sh1}), we recover the stochastic
Keller-Segel equation (\ref{sks14}). This is valid in a strong
friction limit $\xi\rightarrow +\infty$ with $\xi D_{*}\sim 1$. As in
Sec. \ref{sec_mm}, we can obtain a more general model taking into
account some memory effects. Indeed, if we neglect only the nonlinear term
$\nabla (\rho {\bf u}\otimes {\bf u})$ in Eq. (\ref{sh2}), we find
\begin{eqnarray}
\label{sh4}
\chi\frac{\partial}{\partial t}(\rho {\bf u})=-D_* \nabla\rho+\chi\rho\nabla c-\rho {\bf u}-\sqrt{2D_{*}\rho}\ {\bf R}({\bf r},t).
\end{eqnarray}
Taking the time derivative of Eq. (\ref{sh1}) and substituting
Eq. (\ref{sh4}) in the resulting expression, we obtain the
stochastic Cattaneo model of chemotaxis
\begin{eqnarray}
\label{mm4b}
\chi\frac{\partial^{2}\rho}{\partial t^{2}}+\frac{\partial\rho}{\partial t}
=\nabla\cdot  \left (D_*\nabla\rho-\chi\rho\nabla c\right )+\nabla\cdot (\sqrt{2D_* \rho}{\bf R}).
\end{eqnarray}

\section{Stochastic kinetic  models of chemotaxis}
\label{sec_sk}

In order to take into account fluctuations in a rigorous way, we must
start from a microscopic description of the dynamics of the
chemotactic species. In Sec. \ref{sec_sks}, we have considered an
overdamped dynamics. However, according to recent observations in
biology (as discussed in Sec. \ref{sec_mm}), it is important to take
into account the inertia of the particles. A kinetic model of
chemotaxis taking into account finite $N$ effects and inertial effects
has been proposed in Chavanis \& Sire \cite{csbio}. In the simplest
case, the motion of the biological entities is described by $N$
coupled stochastic Langevin equations of the form
\begin{eqnarray}
\label{sk1} \frac{d{\bf r}_i}{dt}={\bf v}_i,
\end{eqnarray}
\begin{eqnarray}
\label{sk2} \frac{d{\bf v}_i}{dt}=-\xi {\bf
v}_i+\nabla c_{d}({\bf r}_{i}(t),t)+\sqrt{2D}{\bf R}_i(t),
\end{eqnarray}
\begin{eqnarray}
\label{sk3}
\frac{\partial c_{d}}{\partial t}=D_{c}\Delta c_{d}-kc_{d}+h\sum_{i=1}^{N}\delta({\bf r}-{\bf r}_{i}(t)),
\end{eqnarray}
where $\xi$ is a friction coefficient and $D$ a diffusion coefficient
in velocity space. We can introduce an effective temperature $T_{eff}$
through the Einstein relation $T_{eff}={D}/{\xi}$ \cite{csbio,gfp}. The
overdamped stochastic equations (\ref{intro9})-(\ref{intro10}) can be
recovered in a strong friction limit $\xi\rightarrow +\infty$,
neglecting the inertial term in Eq. (\ref{sk2}), and writing $\chi=1/\xi$ and
$D_{*}=D/\xi^2$. We now proceed in deriving the exact kinetic equation
satisfied by the distribution function of cells whose dynamics is
described by the coupled stochastic Langevin equations
(\ref{sk1})-(\ref{sk3}). The exact distribution function, expressed in
terms of $\delta$-functions, can be written
\begin{eqnarray}
\label{df0}
f_d({\bf
r},{\bf v},t)=\sum_{i=1}^{N}f_{i}({\bf r},{\bf v},t)=\sum_{i=1}^{N}\delta({\bf r}-{\bf r}_{i}(t))\delta({\bf
v}-{\bf v}_{i}(t)).
\end{eqnarray}
For any function $F({\bf r},{\bf v})$, we have $F({\bf r}_{i}(t),{\bf v}_{i}(t))=\int f_{i}({\bf r},{\bf v},t)F({\bf r},{\bf v})d{\bf r}d{\bf v}$. Now, using Ito's calculus, one has
\begin{eqnarray}
\label{df1}
\frac{dF({\bf r}_{i},{\bf v}_{i})}{dt}=\int f_{i}({\bf r},{\bf v},t)\biggl\lbrack \nabla_{\bf r} F({\bf r},{\bf v})\cdot {\bf v}-\xi\nabla_{\bf v} F({\bf r},{\bf v})\cdot {\bf v}+\nabla_{\bf v} F({\bf r},{\bf v})\cdot\nabla c_{d}({\bf r},t)\nonumber\\
+\sqrt{2D} \nabla_{\bf v} F({\bf r},{\bf v})\cdot {\bf R}_{i}(t)+D\Delta_{\bf v}F({\bf r},{\bf v})\biggr\rbrack d{\bf r}d{\bf v}.
\end{eqnarray}
Integrating by parts, we obtain
\begin{eqnarray}
\label{df2}
\frac{dF({\bf r}_{i},{\bf v}_{i})}{dt}=\int F({\bf r},{\bf v})\biggl\lbrack -{\bf v}\cdot \frac{\partial f_{i}}{\partial {\bf r}}({\bf r},{\bf v},t)+\xi\frac{\partial}{\partial {\bf v}}\cdot (f_{i}({\bf r},{\bf v},t){\bf v})-\nabla c_{d}({\bf r},t)\cdot \frac{\partial f_{i}}{\partial {\bf v}}({\bf r},{\bf v},t)\nonumber\\
-\sqrt{2D} \frac{\partial}{\partial {\bf v}}(f_{i}({\bf r},{\bf v},t){\bf R}_{i}(t))+D\Delta_{\bf v}f_{i}({\bf r},{\bf v},t)\biggr\rbrack d{\bf r}d{\bf v}.
\end{eqnarray}
Then, using $dF({\bf r}_i,{\bf v}_i)/dt=\int \partial_t f_{i}({\bf r},{\bf v},t)F({\bf r},{\bf v}) d{\bf r}d{\bf v}$ and comparing with Eq. (\ref{df2}), we get 
\begin{eqnarray}
\label{df3}
\frac{\partial f_{i}}{\partial t}+{\bf v}\cdot \frac{\partial f_{i}}{\partial {\bf r}}+\nabla c_{d}\cdot \frac{\partial f_{i}}{\partial {\bf v}}=\frac{\partial}{\partial {\bf v}}\cdot \left (D\frac{\partial f_{i}}{\partial {\bf v}}+\xi f_{i}{\bf v}\right )-\sqrt{2D}\frac{\partial}{\partial {\bf v}}\cdot \left (f_{i}{\bf R}_{i}\right ).
\end{eqnarray}
Summing this relation over the $i$, we finally obtain
\begin{eqnarray}
\label{df4}
\frac{\partial f_{d}}{\partial t}+{\bf v}\cdot \frac{\partial f_{d}}{\partial {\bf r}}+\nabla c_{d}\cdot \frac{\partial f_{d}}{\partial {\bf v}}=\frac{\partial}{\partial {\bf v}}\cdot \left (D\frac{\partial f_{d}}{\partial {\bf v}}+\xi f_{d}{\bf v}\right )-\sqrt{2D}\sum_{i=1}^{N}\frac{\partial}{\partial {\bf v}}\cdot \left (f_{i}{\bf R}_{i}\right ).
\end{eqnarray}
Now, proceeding like in \cite{dean}, the last term can be rewritten:
\begin{eqnarray}
\label{df5}
-\sum_{i=1}^{N}\frac{\partial}{\partial {\bf v}}\cdot \left (f_{i}({\bf r},{\bf v},t){\bf R}_{i}(t)\right )=\frac{\partial}{\partial {\bf v}}\cdot (f_{d}^{1/2}({\bf r},{\bf v},t){\bf Q}({\bf r},{\bf v},t)),
\end{eqnarray} 
where ${\bf Q}({\bf r},{\bf v},t)$ is a Gaussian random field such
that $\langle {\bf Q}({\bf r},{\bf v},t)\rangle={\bf 0}$ and $\langle
Q_{\alpha}({\bf r},{\bf v},t)Q_{\beta}({\bf r}',{\bf v}',t')\rangle\\
=\delta_{\alpha\beta}\delta({\bf r}-{\bf r}')\delta({\bf v}-{\bf
v}')\delta(t-t')$. Therefore, the system of equations satisfied by the exact distribution function expressed
in terms of $\delta$-functions is
\begin{eqnarray}
\label{sk4}
\frac{\partial f_d}{\partial t}+{\bf v}\cdot \frac{\partial f_d}{\partial {\bf r}}+\nabla c_d\cdot \frac{\partial f_d}{\partial {\bf v}}=\frac{\partial}{\partial {\bf v}}\cdot \left ( D\frac{\partial f_d}{\partial {\bf v}}+\xi  f_d {\bf v}\right )+\frac{\partial}{\partial {\bf v}}\cdot \left (\sqrt{2Df_d}{\bf Q}({\bf r},{\bf v},t)\right ),
\end{eqnarray}
\begin{eqnarray}
\label{sk5}
\frac{\partial c_{d}}{\partial t}=D_{c}\Delta c_{d}-kc_{d}+h\int f_{d}({\bf r},{\bf v},t) d{\bf v}.
\end{eqnarray}
This will be called the stochastic Kramers equation
of chemotaxis for the exact distribution function.  Using
Eq. (\ref{sks8}), it can be written
\begin{eqnarray}
\label{sk6w}
\frac{\partial f_d}{\partial t}+{\bf v}\cdot \frac{\partial f_d}{\partial {\bf r}}+h \int d{\bf r}'d{\bf v}'\int_{0}^{t}dt' \nabla G({\bf r}-{\bf r}',t-t') f_{d}({\bf r}',{\bf v}',t')\cdot \frac{\partial f_{d}}{\partial {\bf v}}({\bf r},{\bf v},t)\nonumber\\
=\frac{\partial}{\partial {\bf v}}\cdot \left ( D\frac{\partial f}{\partial {\bf v}}+\xi  f{\bf v}\right )+\frac{\partial}{\partial {\bf v}}\cdot \left (\sqrt{2Df_d}{\bf Q}({\bf r},{\bf v},t)\right ).
\end{eqnarray}
If we average over the noise and introduce the smooth distribution
function $f({\bf r},{\bf v},t)=\langle f_{d}({\bf r},{\bf
v},t)\rangle$, we recover Eq. (60) of Chavanis \& Sire \cite{csbio}:
\begin{eqnarray}
\label{sk6}
\frac{\partial f}{\partial t}+{\bf v}\cdot \frac{\partial f}{\partial {\bf r}}+h\frac{\partial}{\partial {\bf v}}\cdot \int d{\bf r}'d{\bf v}'\int_{0}^{t}dt' \nabla G({\bf r}-{\bf r}',t-t')\langle f_{d}({\bf r},{\bf v},t)f_{d}({\bf r}',{\bf v}',t')\rangle \nonumber\\
=\frac{\partial}{\partial {\bf v}}\cdot \left ( D\frac{\partial f}{\partial {\bf v}}+\xi  f{\bf v}\right ).
\end{eqnarray}
If we make a mean field approximation $\langle f_{d}({\bf r},{\bf
v},t)f_{d}({\bf r}',{\bf v}',t')\rangle\simeq f({\bf r},{\bf v},t)f({\bf
r}',{\bf v}',t')$, we recover Eqs. (66)-(68) of Chavanis \& Sire
\cite{csbio}:
\begin{eqnarray}
\label{sk7}
\frac{\partial f}{\partial t}+{\bf v}\cdot \frac{\partial f}{\partial {\bf r}}+\nabla c\cdot \frac{\partial f}{\partial {\bf v}}=\frac{\partial}{\partial {\bf v}}\cdot \left ( D\frac{\partial f}{\partial {\bf v}}+\xi  f {\bf v}\right ),
\end{eqnarray}
\begin{eqnarray}
\label{sk8}
\frac{\partial c}{\partial t}=D_{c}\Delta c-kc+h\int f({\bf r},{\bf v},t) d{\bf v}.
\end{eqnarray}
This can be viewed as a mean field Kramers equation of chemotaxis in
the same way that the Keller-Segel model can be viewed as a
Smoluchowski equation of chemotaxis. In fact, the Keller-Segel model
(\ref{intro1})-(\ref{intro2}) can be recovered from
Eqs. (\ref{sk7})-(\ref{sk8}) in a strong friction limit
$\xi\rightarrow +\infty$ by using a Chapman-Enskog expansion
\cite{lemou} or a method of moments \cite{csbio}. Let us note, for
future reference, that the steady solutions of the  mean field Kramers equation of chemotaxis  correspond
to a mean field Maxwell-Boltzmann-like distribution
\begin{eqnarray}
\label{skt}
f=A' e^{-\beta(v^2/2-c)},
\end{eqnarray}
where $\beta=1/T_{eff}$ is the inverse effective temperature. If we
integrate this distribution over the velocitities we recover the
distribution (\ref{sks13b}) that is the steady solution of the
Keller-Segel model (\ref{intro1})-(\ref{intro2}).

As discussed in the Introduction, the mean field approximation may not
always give a good description of the dynamics. On the other hand,
Eqs. (\ref{sk4})-(\ref{sk5}) for the distribution function expressed
in terms of $\delta$-functions are exact but they are  too complicated for
practical purposes because they  contain exactly the same information as
the $N$-body stochastic Langevin equations (\ref{sk1})-(\ref{sk3}).
Therefore, as in Sec. \ref{sec_sks}, we shall introduce a simplified
kinetic equation for a coarse-grained distribution function
$\overline{f}({\bf r},{\bf v},t)$ which smoothes out the exact
distribution function ${f}_d({\bf r},{\bf v},t)$ while keeping track
of fluctuations.  We propose the simplified stochastic model
\begin{eqnarray}
\label{sk9}
\frac{\partial \overline{f} }{\partial t}+{\bf v}\cdot \frac{\partial \overline{f} }{\partial {\bf r}}+\nabla \overline{c}\cdot \frac{\partial \overline{f} }{\partial {\bf v}}=\frac{\partial}{\partial {\bf v}}\cdot \left ( D\frac{\partial \overline{f} }{\partial {\bf v}}+\xi  \overline{f}  {\bf v}\right )+\frac{\partial}{\partial {\bf v}}\cdot \left (\sqrt{2D\overline{f}}{\bf Q}({\bf r},{\bf v},t)\right ),
\end{eqnarray}
\begin{eqnarray}
\label{sk10}
\frac{\partial \overline{c}}{\partial t}=D_{c}\Delta \overline{c}-k\overline{c}+h\int \overline{f}({\bf r},{\bf v},t) d{\bf v}.
\end{eqnarray}
This model takes into account inertial effects and fluctuations so
that it should provide a good description of the dynamics of
chemotactic species. As shown in Appendix B of \cite{hb5}, the form of
the noise in these equations can be obtained by applying the general
theory of fluctuations developed by Landau \& Lifshitz \cite{ll}.
 
Let us try to make a connexion with the hydrodynamic equations
introduced phenomenologically in Sec. \ref{sec_sh}. Taking the
hydrodynamic moments of the stochastic Kramers equation (\ref{sk9})
and proceeding as in
\cite{csbio}, we obtain
\begin{eqnarray}
\label{sk11}
\frac{\partial\rho}{\partial t}+\nabla\cdot (\rho {\bf u})=0,
\end{eqnarray}
\begin{eqnarray}
\label{sk12}
\frac{\partial}{\partial t}(\rho u_{i})+\frac{\partial}{\partial x_{j}}(\rho u_{i}u_{j})=-\frac{\partial P_{ij}}{\partial x_{j}}+\rho\frac{\partial c}{\partial x_{i}}-\xi\rho u_{i}-\int \sqrt{2Df}Q_{i}d{\bf v}, 
\end{eqnarray}
where $\rho({\bf r},t)=\int f d{\bf v}$ is the density, ${\bf u}({\bf r},t)=(1/\rho)\int f{\bf v}d{\bf v}$ is the local velocity, ${\bf w}={\bf v}-{\bf u}({\bf r},t)$ is the relative velocity and $P_{ij}=\int f w_{i}w_{j}d{\bf v}$ is the pressure tensor. Defining ${\bf g}({\bf r},t)\equiv \int \sqrt{2Df}{\bf Q}d{\bf v}$, it is clear that ${\bf g}$ is a Gaussian noise and that its correlation function is
\begin{eqnarray}
\label{sk13}
\langle g_{i}({\bf r},t)g_{j}({\bf r}',t')\rangle=2D\int \sqrt{f({\bf r},{\bf v},t)f({\bf r}',{\bf v}',t')}\langle Q_{i}({\bf r},{\bf v},t)Q_{j}({\bf r}',{\bf v}',t')\rangle d{\bf v}d{\bf v}'\nonumber\\
=2D\delta_{ij}\delta({\bf r}-{\bf r}')\delta(t-t')\int f({\bf r},{\bf v},t) d{\bf v}=2D\delta_{ij}\delta({\bf r}-{\bf r}')\delta(t-t')\rho({\bf r},t). 
\end{eqnarray}
Therefore, the equation for the momentum (\ref{sk12}) can be rewritten 
\begin{eqnarray}
\label{sk14}
\frac{\partial}{\partial t}(\rho u_{i})+\frac{\partial}{\partial x_{j}}(\rho u_{i}u_{j})=-\frac{\partial P_{ij}}{\partial x_{j}}+\rho\frac{\partial c}{\partial x_{i}}-\xi\rho u_{i}-\sqrt{2D\rho}R_{i}({\bf r},t).
\end{eqnarray}
This equation is not closed since the pressure tensor depends on the
next order moment of the velocity. If, following \cite{csbio}, we make
a local thermodynamic equilibrium (L.T.E.) approximation $f_{LTE}({\bf
r},{\bf v},t)\simeq (\beta /{2\pi})^{d/2}
\rho({\bf r},t) e^{-\beta w^2/2}$ to compute the
pressure tensor, we find that $P_{ij}\simeq
T_{eff}\rho\delta_{ij}$. In that case, Eqs. (\ref{sk11}) and
(\ref{sk14}) return the stochastic damped Euler equations
(\ref{sh1})-(\ref{sh2}). We recall, however, that there is no rigorous
justification for this local thermodynamic equilibrium
approximation. Therefore, it does not appear possible to rigorously
derive the damped hydrodynamic equations (\ref{sh1})-(\ref{sh2}) from
the Kramers equation (\ref{sk9})-(\ref{sk10}) by a systematic
procedure. Alternatively, if we consider the strong friction limit
$\xi\rightarrow +\infty$ for fixed $\beta$, implying
$D=\xi/\beta\rightarrow +\infty$, the first term in the r.h.s. of
Eq. (\ref{sk9}) implies that $f({\bf r},{\bf v},t)\simeq
(\beta/{2\pi})^{d/2}
\rho({\bf r},t) e^{-\beta v^2/2}+O(1/\xi)$, ${\bf u}=O(1/\xi)$ and $P_{ij}=T_{eff}\rho\delta_{ij}+O(1/\xi)$ \cite{csbio}. To leading order in $1/\xi$, Eq. (\ref{sk14}) becomes
\begin{eqnarray}
\label{sk15}
\rho {\bf u}\simeq -\frac{1}{\xi}\left (T_{eff}\nabla\rho-\rho\nabla c+\sqrt{2D\rho}{\bf R}({\bf r},t)\right ).
\end{eqnarray}
Inserting Eq. (\ref{sk15}) in the continuity equation (\ref{sk11}) and
recalling that $T_{eff}=D/\xi=\xi D_*$ and $\chi=1/\xi$, we recover
the stochastic Keller-Segel model (\ref{sks14})-(\ref{sks15}).  It is
therefore possible to rigorously derive the stochastic Keller-Segel
model (\ref{sks14})-(\ref{sks15}) from the stochastic Kramers equation
(\ref{sk9})-(\ref{sk10}) in the strong friction limit $\xi\rightarrow
+\infty$.

\section{Conclusion}
\label{sec_conclusion}

In this paper, we have derived generalized Keller-Segel models of
chemotaxis taking into account fluctuations. This leads to stochastic
kinetic equations instead of deterministic equations. Fluctuations
become important close to a critical point \cite{kadanoff,grima,hb5},
so it is valuable to have a model of chemotaxis going beyond the mean
field approximation and taking into account fluctuations. The
divergence of the spatial correlation function close to the critical
point has been analyzed in detail in
\cite{hb5} for Brownian particles interacting through a binary potential. These particles are described by a stochastic Smoluchowski equation coupled to
the markovian field equation (\ref{intro8}). The general methods
developed in
\cite{hb5} can be extended to the stochastic Keller-Segel model (\ref{sks14}) 
coupled to the non-Markovian field equation (\ref{intro2}). Accounting
for fluctuations is also important when the number of particles $N$ is
small and when there exists several metastable states.  In that case,
fluctuations can trigger dynamical phase transitions from one state to
the other.

We have also introduced kinetic models of chemotaxis in phase space
taking into account inertial effects. In the strong friction limit, we
recover the Keller-Segel model describing an overdamped dynamics. We
have discussed the relation between the kinetic equations in phase
space and the hydrodynamic equations introduced
phenomenologically. Finally, we have shown how the Cattaneo model of
chemotaxis \cite{dh} could be obtained from these hydrodynamic
equations.

This paper and \cite{hb5} are the first attempts to include
fluctuations in the kinetic equations of chemotaxis (the main results
were given in \cite{hb5} and they have been discussed here
specifically with more details and amplification). In view of the
importance of the Keller-Segel model in biology, the stochastic
equations that we propose can have a lot of applications and can open
the way to many new investigations. Their detailed numerical and
analytical study is therefore of considerable interest. We hope to
come to these problems in future works.

\vskip1cm

{\it Note added:} Until now, fluctuations have been ignored by people
working on chemotaxis. Therefore, Ref. \cite{hb5} and the present
paper are the first attempts to include fluctuations in the
Keller-Segel model. However, after submission of these papers, a paper
by Tailleur \& Cates [arXiv:0803.1069] (now published as
Phys. Rev. Lett. {\bf 100}, 218103 (2008)) came out on a related
subject. These authors also consider the effect of fluctuations in the
motion of bacteria. However, their goal is different. They are
mainly interested in deriving transport coefficients from microscopic models, so they  do not take into account the long-range interaction between
bacteria due to chemotaxis. Alternatively, in our approach, the
transport coefficients $D_*$ and $\chi$ appearing in the Langevin
equations are introduced phenomenologically but long-range interaction
between bacteria due to chemotaxis is fully taken into
account. Therefore, these two independent studies are complementary to
each other.

\end{document}